# An Analysis of Early-Stage Functional Safety Analysis Methods and Their Integration into Model-Based Systems Engineering

*Jannatul Shefa[1] and Taylan G. Topcu[2]*

[1]Grado Department of Industrial and Systems Engineering, Virginia Tech, Blacksburg, VA, USA; shefa@vt.edu

[2]Grado Department of Industrial and Systems Engineering, Virginia Tech, Blacksburg, VA, USA; ttopcu@vt.edu

**Abstract**

*As systems become increasingly complex, conducting effective safety analysis in the earlier phases of a systems' lifecycle is essential to identify and mitigate risks before they escalate. To that end, this paper investigates the capabilities of key safety analysis techniques, namely: Failure Mode and Effects Analysis (FMEA), Functional Hazard Analysis (FHA), and Functional Failure Identification and Propagation (FFIP); along with the current state of the literature in terms of their integration into Model-Based Systems Engineering (MBSE). A two-phase approach is adopted. The first phase is focused on contrasting FMEA, FHA, and FFIP techniques, examining their procedures, along with a documentation of their relative strengths and limitations. Our analysis highlights FFIP's capability in identifying emergent system behaviors, second order effects, and fault propagation; thus, suggesting it is better suited for the safety needs of modern interconnected systems. Second, we review the existing research on the efforts to integrate each of these methods into MBSE. We find that MBSE integration efforts primarily focus on FMEA and integration of FHA and FFIP is nascent. Additionally, FMEA-MBSE integration efforts could be organized in four: model-to-model transformation, use of external customized algorithms, built-in MBSE packages, and manual use of standard MBSE diagrams. While our findings indicate a variety of MBSE integration approaches, there is no universally established framework or standard. This leaves room for an integration approach that could support the ongoing Digital Engineering transformation efforts by enabling a more synergistic lifecycle safety management methods and tools.*

## 1. Introduction

Today's engineered systems are complex [1] with increasing demands for cyber-physical properties [2], interconnectedness, and intelligence [3]; which in turns leads to a large number of components, functions, interfaces and branching paths. This increases the possibility of multiple interrelated failure events that could occur jointly or sequentially, leading to system failures and accidents that are difficult to predict *a priori* [4]. These trends are exacerbating existing challenges complex system development, and call for increased coherence in system lifecycle management [5] [6]. Failure to do this effectively and efficiently leads to undesirable outcomes that are commonly manifested in design rework, cost overruns, or catastrophic accidents after deployment [7, 8]. Modern safety analysis (SA) efforts must account for these factors to ensure comprehensive risk mitigation.

To manage these risks proactively, SA methods need to be integrated as early in the system development process as possible. This enables identification of hazards and failure scenarios before the operational and maintenance lifecycle begins; and design decisions can be made to enhance system safety while minimizing the cost of mitigation efforts [9]. However, there is limited information regarding the physical composition of the system during early design phases [10]; thus, functional SA methods are more appropriate. A plethora of functional SA techniques have been developed for this purpose. These include qualitative, quantitative, or mixed-methods approaches. Common SA techniques include the fault tree analysis (FTA) [11], FHA [12], FMEA [13], Markov analysis (MA) [14], dependence diagrams (DD) [15], FFIP [9], and Probabilistic Risk Assessment (PRA) [16]. However, these methods are often performed independently with separate tools [17] and there are uncertainties regarding which SA method is best suited for a given system or a specific situation [18].

Additionally, there is an ongoing push in the systems engineering (SE) community for digital transformation [19], a paradigm shift enabled by advancements in information technology [20] that aims to move system lifecycle management approaches beyond the traditional document-based methods. Digital transformation builds off from the recent efforts of Model-Based Systems Engineering (MBSE) [21]. MBSE has emerged as a leading approach systems development, with hopes of unifying various modeling perspectives, tools, and analysis methods within a single model repository. This repository serves as the foundation for all development and analysis efforts across the system lifecycle [20], fulfilling the needs of both systems and safety engineers while minimizing outdated or duplicated documentation [22].

However, similar to the disciplinary resistance and cognitive entrenchment experienced in MBSE adoption, digital transformation also calls for a change in traditional SA procedures [21, 23], which were typically performed in a mutually exclusive manner, usually through manual extraction of information from system models—a process prone to errors and time delays. Naturally, as designs evolve, these mutually exclusive safety studies often relied on outdated design information, leading to inconsistencies and ultimately, rework [24]. By integrating SA synergistically within MBSE, configuration management can be done more efficiently, while maintaining consistency and allowing safety and design teams to collaborate seamlessly within a shared digital ecosystem.

To that end, the objective of this study is to compare three prominent early-stage functional safety models, namely FMEA, FHA, and FFIP, and discuss how these methods are currently being integrated into MBSE. More specifically, we pursue the following research questions:

*RQ1:* How do the capabilities of these early-stage functional safety methods compare?

*RQ2:* How are these methods being integrated into MBSE in current practice?

This study synthesizes the capabilities of the aforementioned three SA techniques in addressing key safety concerns of complex systems, including emergent behavior, formalized risk assessment and ranking, and the evaluation of multiple failure interactions. The findings also indicate that FMEA-MBSE integration has received significantly more attention in the literature compared to FHA and FFIP.

## 2. Background

*2.1 Model-based systems engineering and product lifecycle management*

Product Life Cycle Management (PLM) can be viewed as a business strategy aimed at managing data, information, knowledge, and expertise critical for establishing and maintaining a product-focused knowledge environment across all phases of a system's lifecycle [24, 25]. An integrated PLM environment facilitates collaboration among informed decision-makers by merging and incorporating the perspectives of different stakeholders throughout the product's lifecycle. Over the past decade, PLM vendors have proactively incorporated MBSE capabilities into their solutions, starting with the connection between the system design and detailed design phases [5].

MBSE formalizes the practice of SE by developing and utilizing a unified model that is consistent with different domain-specific product models and remains consistent throughout the entire system life cycle [26]. In the MBSE framework, an integrated and coherent system model serves as a central repository for product data relationships and development decisions. This contrasts with document-based traditional SE approaches, where individual design artifacts are managed separately, often leading to disjointed references that can fall out of sync or be disconnected. In such environments, a significant portion of lifecycle costs is associated with the development and maintenance of these fragmented artifacts [27].

The significance of traceability and reproducibility in models and simulations, a key aspect in developing increasingly complex products, necessitates the seamless integration of MBSE and PLM. As the role of decisions based on modeling and simulation expands, INCOSE introduced the term "Digital Engineering" to describe MBSE enhanced by simulation technologies. Digital Engineering promotes an integrated model-based approach by leveraging digital methods, tools, processes, and artifacts [5]. By allowing the system model to be updated in real-time, organizations can avoid working with outdated information. Additionally, utilizing a common model promotes the early consideration of safety requirements during the conceptual design phase, potentially reducing the number of iterations and design changes later in the development process [24, 25]. The integration of SA data generation within the MBSE environment enhances task execution efficiency. This approach allows for the integration of various engineering modeling data and analyses, significantly reducing the need for manual efforts in locating relevant design and product development information.

The lack of traceability among system functions, failure conditions, and safety requirements increases the risk of inconsistencies and costly errors [28]. Thus, seamlessly integrating SA techniques (both early-stage and late-stage after deployment) with MBSE has the capability to enhance traceability and consistency between both design and safety aspects [29]. As organizations increasingly adopt model-based approaches, system models can change rapidly, often outpacing SA efforts. Safety engineers are required to manually extract data from the evolving system model for their analyses, meaning the resulting safety assessments are only accurate for the specific iteration of the system. Any modifications to the design render previous safety analyses obsolete, leading to wasted time and resources [24].

*2.2. Early-stage SA methods*

Early-stage SA allows systematically examining systems in order to identify potential hazards and safety characteristics [30]. Incorporating SA early in the system lifecycle can help in avoiding late-stage design changes can be significantly more costly. To put things in reference, fault correction efforts roughly cost six times more in the conceptual design phase [31], one hundred times more during the development phase, and a thousand times more expensive in the production or testing phase [25]. This highlights the importance of integrating safety assessments early in the design process to ensure timely validation of safety requirements. Out of the existing early-stage SA methods, we present a comparative analysis of the capabilities of widely used FMEA, FHA, and relatively more recent FFIP methods.

FMEA is a qualitative and inductive method aimed at identifying potential issues early in the design phase of a system or product that could impact safety and performance. This bottom-up analysis method focuses on implementing countermeasures to reduce or eliminate the effects of the identified failure modes [18]. FHA is also qualitative and inductive in nature and focuses on system functions. This method pinpoints system functions and assesses the impact of malfunctions [29]. Finally, FFIP is a novel approach to designing complex reliable systems. It combines function, structure, and behavior modeling to simulate how failures propagate and lead to functional breakdowns, to identify mitigation strategies. This approach integrates hierarchical system models with behavioral simulations and qualitative reasoning to better understand failure pathways [9].

While there has been a considerable amount of work comparing FMEA and FHA, and relatively less research on their integration with MBSE; there are even fewer studies that focus on FFIP within the context of MBSE. To that end, we studied the current studies in the literature to understand the current state of research on integrating these three SA methods into MBSE. Furthermore, this study is inspired by the work of Van Bossuyt et al. [4] that compared SA methods focusing on their capability to identify failure propagation paths, quantify failure probability outcomes, and detect irrationality initiators in the system. However, our study departs from their work by expanding the scope of comparison and characterization of the current practices to integrate these SA methods with MBSE.

## 3. Research Design

This study adopts a two-phase approach to compare the key steps and capabilities of FMEA, FHA, and FFIP followed by exploring their integration into MBSE for enhanced SA. In the first phase, we conducted a literature review to evaluate the risk analysis capabilities offered by FMEA, FHA, and FFIP and synthesized our findings. The focus of this analysis was to understand these individual techniques, examining the steps they execute to identify, evaluate, and mitigate risks. Particular attention was given to how these steps align or differ across the techniques in terms of scope, capability, and process structure. In the second phase, existing efforts in the literature to integrate these three failure analysis techniques were explored, either independently or in combination. This phase involved reviewing studies that focused on integrating MBSE with at least one of FMEA, FHA, or FFIP. The goal was to understand the current state of research on combining these methods to create a unified framework that supports holistic failure analysis.

## 4. Description of the SA techniques

4.1. The generic FHA procedure

While different sources may present variations in wording and the number of steps involved in FHA, the core process generally includes the following steps as outlined by [32]:
    (1) The system is described using information from the conceptual design phase. This step involves gathering and interpreting the system's functional architecture and modeling how the functions interact, aiming to create a functional hierarchy, function flow block diagrams, and a function matrix for the system [33].
    (2) Hazards are identified for each function [33] or by examining combinations of system functions, operational modes, and potential functional failure modes.
    (3) After identifying hazards, the potential consequences of each hazard are determined.
    (4) The next step is to analyze possible causes or scenarios that could lead to hazards. Causal factors may involve elements from the system's conceptual design, function model, or past operational experiences.
    (5) The risks associated with each hazardous event are evaluated based on the severity of consequences and the likelihood of each causal scenario occurring. The risk assessment in hazard analysis can be either qualitative (using scales) or semi-quantitative, e.g. assigning risk priority numbers (RPN) [34].
    (6) Finally, the analysis may result in recommendations to prevent or mitigate identified hazards; or to generate functional requirements to guide detailed system design.

4.2. A generic FMEA procedure

The key stages of FMEA are outlined below:
    (1) First, the system's functional requirements are identified and a detailed breakdown of these functions is developed.

(2) Next the failure modes (FM) of components or functions are established. An FM refers to any means by which a component or process step could fail to carry out its intended function(s) (e.g., a bearing may crack or break).
(3) This step involves pinpointing the part-level causes that could lead to the FMs identified in Step 2.
(4) The effects of failure are analyzed at both local and higher system levels where applicable.
(5) Next, the severity (S) score is assigned to each FM's effect, reflecting its impact. Occurrence (O) and detectability (D) scores are given to represent the likelihood of the FM happening and the difficulty of detecting the cause or FM, respectively. The RPN is then calculated by multiplying S, O, and D to determine the risk level of each FM [35].
(6) Lastly, actions are proposed to mitigate or compensate for the risks identified [36].
Fig. 1 illustrates the steps of the FMEA and FHA procedure as discussed above.

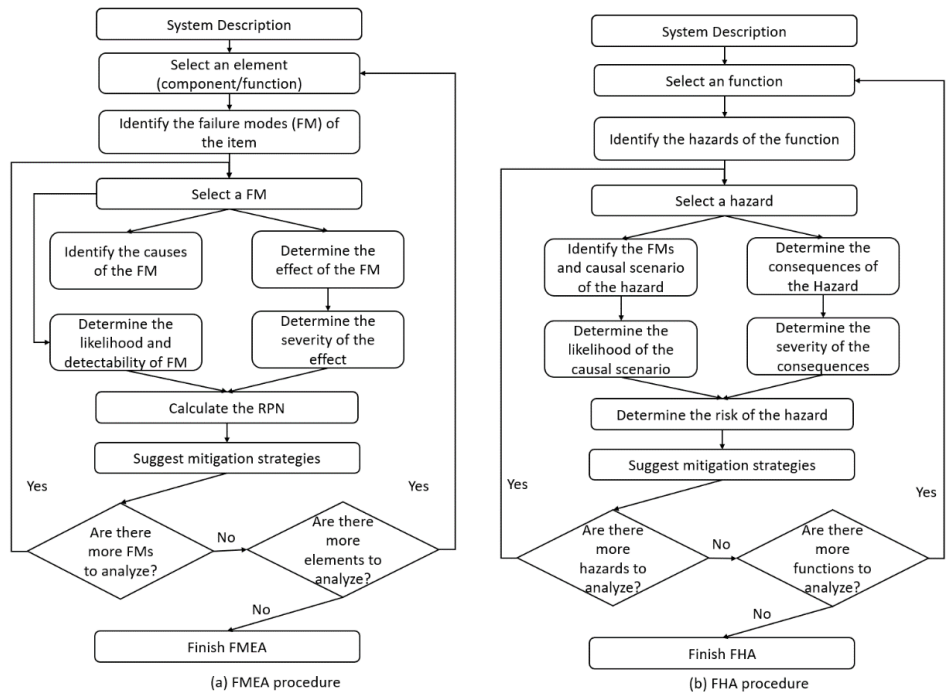

Fig. 1. Generic FMEA (adapted from [37]) and FHA procedures

4.3. A generic FFIP procedure

The key stages of FFIP, adapted from [38], are outlined below:
(1) The first step involves Mapping the Function to Component. The functional model or function flow block diagram illustrates how the system's intended functional transformations occur. A configuration flow graph is utilized to show system components and paths through which interaction terms of energy, material, and signals are transferred.
(2) In most FFIP research, component behaviors are modeled using a discrete state-based approach. Each component is assigned multiple nominal and failure states. Behaviors in these states can be represented as continuous values (e.g., current drawn from a battery) or discrete levels (such as low, high, or zero).
(3) FFIP assumes that the system's functions are intended to perform actions on flows of energy, material, or signals, allowing the identification of a "state" for each function based on unexpected behaviors. The possible states of a function could be classified based on the designer's intent as follows: (i) Healthy—The function operates as intended or is idle; (ii) Degraded—The function works but not at optimal levels; (iii) Lost—The function fails completely, meaning the flow is not processed; and (iv) Lost Recoverable/No Flow—The system is in a faulty state, preventing any flow from being processed. These function states may differ based on the designer's choice. The function failure logic (FFL) reasoner is used to determine the state of a function during the simulation in an event scenario.
(4) Finally, the fault scenarios are simulated. The ultimate goal of modeling behavior, connectivity, and functions is to assess how the system reacts to potential faults. The scenario simulation uses the behavioral and functional failure logic to evaluate the system's response to discrete failures and to reason about potential failures.
Fig. 2 illustrates the key steps of the FFIP procedure as discussed above.

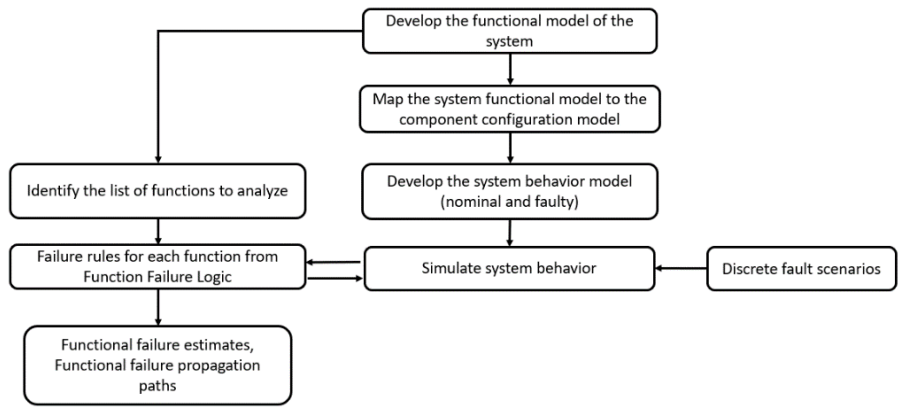

Fig. 2. A generic FFIP procedure

## 5. Findings

In section 5.1, we compare FMEA, FHA, and FPIP capabilities along the eight criteria as listed in Table 1. We also discuss how these three SA techniques are being integrated into MBSE in current practice in section 5.2.

5.1. Comparison of SA techniques

In this section, a comparative analysis of FFIP, FMEA, and FHA is presented. Table 1 summarizes the key capabilities of these methods and each of these characteristics are discussed below.

| Capability | FFIP | FMEA | FHA |
|---|---|---|---|
| Identification of Emergent system behavior | √ | | |
| Formalized risk magnitude assessment and ranking | | √ | √ |
| Failure flow identification and Fault propagation | √ | | |
| Evaluating the Likelihood of occurrence of failure events | | √ | √ |
| Effect of multiple failures or component interactions | √ | | |
| The severity of the event | | √ | √ |
| Inclusion of faulty behavior | √ | | |
| Dependence on analyzer's subjective judgement | | √ | √ |

Table 1. Comparison between SA techniques.

### 5.1.1. Identification of emergent system behaviors

The identification of emergent system behavior is a critical challenge in modern complex systems, where failures often result from the interaction of multiple events rather than predictable, isolated component failures [4]. The FFIP approach can address such behaviors by supporting design decisions and accounting for system failures arising from complex interactions, rather than individual component failures. This can be achieved by simulating fault propagation beyond nominal component connections and applying techniques to identify system-level behaviors simulating lower-level component behaviors [38]. In contrast, traditional FMEA and FHA focus on individual component failures or known hazards, making them ill-suited for identifying system-level failures arising from unforeseen interactions between components [4].

### 5.1.2. Formalized risk magnitude assessment and ranking

FMEA quantifies risk using an RPN. The more the value of RPN, the greater the risk is. Similarly, in FHA, risks for hazardous events are determined by assessing the likelihood of the causal scenarios and the magnitude of the consequences of the hazardous events. This assessment methods also facilitate the magnitude of the associated risk. In contrast, FFIP does not incorporate any formalized way to assess the magnitude of risk associated with the failure events. Thus, quantitative prioritized list is not a direct output of FFIP.

### 5.1.3. Failure flow identification and fault propagation

In the traditional FFIP formulation, fault propagation was traced along the intended flows between system components [10]. FFIP systematically tracks f potential function failures, and follows the failure flow until it exits the system as an output. This approach enables users to visualize how failure states propagate throughout the system and assess whether they present significant risks based on their outcomes [9]. However, real-world systems often experience faults where propagation occurs despite failing components being nominally uncoupled (or directly connected). For example, in a power grid, different substations are designed to operate independently in separate regions. When one substation shuts down, the load shifts to neighboring substations, causing them to overload and fail, leading to widespread blackouts across nominally uncoupled areas. Such observation has led FFIP research to focus on identifying and simulating non-nominal connections, which contribute to emergent failure behaviors [38]. These subsystems detect emerging failures early enough for corrective action. Recent advancements in FFIP have extended its capabilities to evaluate failure flows that bypass nominal pathways and introduced a functional Bayesian approach for developing prognostic and health monitoring subsystems [4, 10, 39]. In contrast, FMEA and FHA lack the ability to model and analyze non-nominal fault propagation and emergent failure behaviors, limiting their effectiveness in complex systems.

*5.1.4. Effects and likelihood of failure*

The FFIP framework serves as a tool for modeling and evaluating the potential effects of failures in the early design stage of a system. As its name suggests, FFIP primarily concentrates on the functional impacts of faults at the component, environmental, and system levels [38]. Among the tools developed under the FFIP methodology, each individual function's response to all possible failure flows is modeled. These responses can include (a) decreasing, increasing, or halting nominal flows leaving the function, (b) passing of failure flows through the function and continuing along either nominal or non-nominal paths, (c) the failure flow blocked in a function while continuing to operate normally, (d) generating new failure flows, or (e) a combination of these outcomes. However, the traditional (baseline version of ) FFIP does not account for the probability of each possible response or failure event or failure sequence [4]. Both the FMEA and FHA determine the effects and likelihood of failure/hazardous events in the process [27, 32].

*5.1.5. Severity of the event*

FMEA and FHA assess the severity of potential failure effects employing various ways. example, a five-point Likert scale that the failure effect from 1 (no impact) to 5 (catastrophic) can be used in FMEA [18]. A similar ranking technique is used in FHA, with hazard severity levels including catastrophic, hazardous, major, minor, and no safety effect [33]. These severity levels are qualitative and largely rely on the subjective judgment of experts [33, 40, 41]. Unlike FMEA and FHA, FFIP does not involve calculating the severity of failures, instead; it focuses on failure flows and their propagation.

*5.1.6. Effect of multiple failures or component interactions*

Causal analysis for hazards is built on different accident models, which can be classified into three types [42]: simple linear models, complex linear models, and systemic models. In simple linear models, accidents result from a direct sequence of causes, and prevention focuses on eliminating a single cause. Complex linear models account for event dependencies that may affect the accident chain. Systemic models view accidents as the result of intricate interactions between system components, requiring control of the overall system state to avoid unsafe conditions. Historically, simple and complex linear models were applied to FHA and FMEA where the focus was on identifying root causes in simpler systems [32]. For example, in a conveyor belt system, lack of regular maintenance causes excessive wear on the drive belt, which then slips, stopping the conveyor abruptly and creating safety risks. This failure occurs in a simple linear sequence, without interaction or interplay between different components. However, as systems get more complex, especially with electronic components, non-physical interdependencies (e.g., electromagnetic interference) between elements become significant contributors to hazardous events, making it harder to capture them using tools like FHA and FMEA [32, 38]. The FFIP approach addresses this challenge by modeling and simulating the propagation of failure flows through complex systems, helping to identify and manage emergent failures caused by interdependencies between system components.

*5.1.7. Inclusion of faulty behavior*

The FFIP framework distinguishes itself from FMEA and FHA primarily through its integration of component-specific faulty behaviors and its reasoning about function failures, which link component actions to functional states [38]. While FMEA focuses on identifying potential failure modes of individual components and their effects on the overall system, it often overlooks the specific behaviors of these components when they fail, resulting in a more generic understanding of risks and a reliance on expert judgment that can introduce subjectivity. Similarly, FHA evaluates hazards based on system functions but lacks a systematic analysis of how individual component failures propagate through functional pathways. In contrast, FFIP enhances this analysis by incorporating detailed, component-specific faulty behaviors that directly inform how failures impact system functionality. By modeling the interactions between component behaviors and functional states, FFIP enables a more accurate identification of potential emergent failures arising from complex interdependencies.

*5.1.8. Subjectivity of analysis*

FMEA systematically evaluates individual components and their failure modes to assess risk and reliability. However, FMEA does not explicitly capture component interactions and relies heavily on subjective expert judgment for assessing failure consequences, which often leads to high variability in analysis results due to differences in expert beliefs [9]. The FHA also is a qualitative, inductive method based on systems functions; that suffer from subjectivity [29]. In contrast, FFIP addresses these issues by integrating hierarchical models of function, configuration, and behavior, combining behavior rules, failure scenarios, FFL analyses, and failure simulations within a unified framework. This integration allows for precise computation of component interactions and resulting failure identifiers, reducing dependency on expert input and making FFIP less subjective compared to FMEA and FHA [40].

### 5.2. Efforts to integrate SA techniques with MBSE

We categorized the efforts to integrate the FMEA, FHA, and FFIP techniques into MBSE into four categories: model-to-model transformation, use of external customized algorithm, built-in MBSE packages, and manual use of standard MBSE diagrams. While we investigated how all three SA methods are integrated into MBSE, we found that FMEA-MBSE integration has drawn significant attention in the literature, and the research on FHA and FFIP integration into MBSE is nascent. Figure 3 highlights the mechanisms and overlaps among the categories. The boundaries of each category are represented by dashed lines in green, red, blue, and orange, respectively.

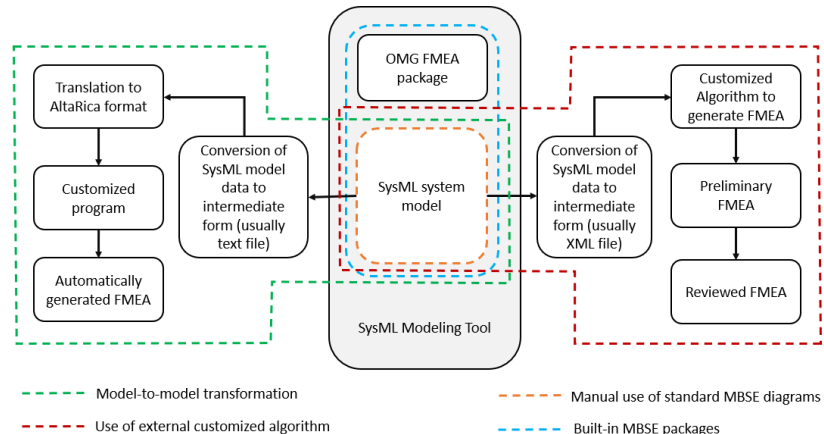

*Fig. 3. Categories to integrate MBSE with FMEA, FHA and FFIP*

In model-to-model transformation, a system model is converted into a format that can be analyzed by existing SA tools, e.g. Altarica [43], Figaro[44], enabling automatic generation of FTAs and FMEAs [28]. For example, AltaRica language is purposefully created to specify the behavior of systems under failure. As the safety model in AltaRica and the design model of a system in MBSE is not identical to each other, either some model-to-model transformation or manual human effort is required to ensure consistency between these two models [45]. Several automated transformation approaches have been proposed for this goal. For instance, Hecht et al. [46] proposed a method to automatically generate FMEA by analyzing various SysML diagrams, such as block definition diagrams (BDDs), internal block diagrams (IBDs), state machines (STMs), and activity diagrams (ADs). The FMEA generator software developed by the authors converts these SysML models into AltaRica by using AltaRica's formalism. Since SysML tools like Cameo Systems Modeler do not natively support AltaRica, model data is exported as a text file and translated into the AltaRica format. The translated model is then processed by a customized program, the Traversal Driver, which triggers the AltaRica engine to simulate state transitions and produce FMEA results. In this method, the generation of FMEA is automated and does not require any further human intervention.

Unlike model-to-model transformation, the second category uses customized algorithms developed during research to generate FMEA, FHA and FMEA artifacts. This allows safety data to be incorporated into the MBSE environment. For example, David et.al. [47] proposed a methodology to automate the creation of a preliminary FMEA report based on functional behaviors in a SysML model followed by a manual review by safety engineers finalizes the FMEA report. This process uses an XML algorithm to analyze and structure data from the model by identifying elements and their connections. The information is then organized into lists to build the FMEA report. The SysML model's structure is mapped with the AltaRica language to allow external commercial tools to generate reliability metrics that can be used on the failure modes in the previous step. A subset of research work in this category uses the SysML modeling language extension approaches to integrate safety-related concepts directly into an existing modeling framework, such as SysML [28]. Mhenni et.al. [24] introduced a method to extend SysML by adding new stereotypes and attributes to store safety-related information such as failure modes, failure rates, severity, causal factors, and failed states. These elements are used to generate safety artifacts. Their method allows SysML models to be exported as XMI (XML Metadata Interchange) files, which are processed by a Python tool designed to extract safety data. This data is then used by the tool to automatically generate preliminary FMEAs and FTAs. The preliminary FMEA is then updated by safety experts and complete FMEA is updated in the system by the custom-built safety tool developed by the authors.

The third approach to integrating the safety and reliability aspect into SysML is to use the package within the Safety and Reliability Analysis Profile for UML submission. Currently, the OMG Safety and Reliability Analysis Profile offers packages for FMEA and there is no existing built-in package for FHA and FFIP. Biggs et al. [48] discussed the contents of the FMEA package and demonstrated its use in SysML through a hybrid vehicle case study. The main concept of the FMEA package is the "FMEAItem" which represents an item of concern for the reliability analysis. Each FMEAItem is related to an item in the system model and linked to failure mode, cause of failure, local and final effect of failure, and prevention and detection control, all of which are linked to the SysML model. The FMEAItem can be updated with severity, occurrence, and/or detectability numbers and calculated RPN [48].

The fourth approach is the most labor-intensive one as it involves conducting failure analysis by manually embedding safety and/or failure information in standard MBSE diagrams. Hence, there is no automation and also, and traditional FMEA or FHA tables are not generated either. For example, Bruno et al. [49] proposed a method that uses STMs in SysML to represent FMEA data, where failure modes identified in the FMEA are depicted as states within the diagram. This method also uses activity diagrams to represent the FHA, capturing the flow of functions in case of failure. The approach shows that Reliability, Availability, Maintainability, and Safety (RAMS) models can be developed using this technique, enabling quicker and more effective assessment of both traditional and innovative on-board system architectures in terms of safety and reliability. Similarly, Clegg et al. [22] developed a SysML safety profile for integrating SA into functional specification documents using IBDs, linking hazards, failure modes, and fault propagation. It demonstrates how fault logic can be modeled using IBDs and traced within FHA, focusing on lower-level functional specifications. Brusa et al. [50] investigated the integration of safety engineering practices in the MBSE environment for aircraft design. They performed the FHA of the aircraft using the information from an AD to manage both the functional and dysfunctional analysis. In short, these are more like different use cases of SysML instead of a distinct approach to conducting an SA. While we predominantly discussed FMEA integration, there are a handful of instances of the use of standard MBSE diagrams to integrate FFIP. Mehrpouyan et al. [51] incorporated the function failure flow of FFIPs into SysML's BDD to assess and compare input and output flows. To support the failure reasoning process, hazard indications resulting from these flows must be displayed within the BDD, ensuring that input and output flow information is available for the failure analysis. Jiao et al. [40] integrated FFIP with SysML to create a unified modeling approach under the Model-based safety analysis (MBSA) framework. This method combines SysML with the FFIP modeling process, helping to build structure, function, and behavior models in FFIP using SysML. By doing this, it addresses the lack of a standardized language for FFIP and simplifies analysis and modification. Additionally, STMs in SysML are enhanced to represent functional failure logic more effectively, extending SysML elements to better capture and analyze failure logic. Nevertheless, this process is rather cumbersome and does not constitute a full-on integration into MBSE.

Table 2. compares the four categories described above. The approaches differ in terms of the information contained in the SysML modeling tools and also in their use of different MBSE diagrams. All four integration approaches target FMEA, in that regard it is the most popular one. We found that so far, only C4 supports FHA and FFIP. Another aspect to emphasize is that C1 and C3 automate SA artifact generation and don't require human supervision; whereas, the other two categories (C2 and C4) need human supervision and has a significant manual labor demand. For example, preliminary FMEA report is revised by safety experts in C2. Furthermore, C1, C2 and C3 require less domain-specific knowledge compared to C4 to understand and interpret the SA once they are generated. C4 requires domain expertise to guide the analysis as the safety information is manually incorporated (or hard-coded) in the MBSE diagrams. Finally, both C1 and C2 depend on customized algorithms/programs external to MBSE tools to perform SA, which may require some additional investment upfront or the creation of these programs. In contrast, C2 and C4 function entirely within the MBSE environment, without the need for external programs.

| | C1: Model-to-model transformation | C2: Use of external customized algorithms | C3: Built-in MBSE packages | C4: Manual use of standard MBSE diagram |
|---|---|---|---|---|
| **SA technique incorporated** | FMEA | FMEA | FMEA | FMEA, FHA, FFIP |
| **Information contained in MBSE modeling tools** | Composition of the system, connections between elements, states of the blocks, propagation and effects from a state to the rest of the block | Requirements definition, system functional architecture, system components and their interaction, safety profile | System Composition, FMEAItem | Functional analysis, system architecture, dysfunctional analysis |
| **MBSE diagrams used** | IBD, BDD, STM, AD | IBD, BDD, AD, Use case diagram | BDD | IBD, BDD, SD, AD |
| **Need for Human supervision generating safety artifacts** | No | Yes | No | Yes |
| **Need for Domain knowledge to interpret safety artifacts** | Less | Less | Less | More |
| **Need for external algorithm/program** | Yes | Yes | No | No |

*Table 2. Comparison between MBSE integration methods.*

## 6. Discussion and Conclusion

As systems grow increasingly complex, effective analysis of system safety becomes more important. To address this, digital engineering tools need to be leveraged to conduct proactive early-stage safety analyses and maintain real-time synchronization of system information across the system lifecycle. To that end, we explored the integration of three key functional SA techniques, FMEA, FHA, and FFIP, within the MBSE environment, along with their comparative strengths and limitations. We present two main findings.

First, in terms of capabilities to cater to the safety needs of modern complex systems, the comparison between FFIP, FMEA, and FHA reveals distinct strengths and limitations. FFIP excels at identifying emergent system behaviors and tracing fault propagation across interconnected components, making it particularly well-suited for capturing failures arising from interactions rather than isolated events. Its' incorporation of component-specific faulty behaviors and failure flow propagation, helps to capture failure dynamics. In contrast, FMEA and FHA prioritize individual component failures and known hazards. FMEA's strength lies in its structured approach to risk quantification, using the RPN to prioritize failure mitigation based on the likelihood, detectability, and severity of failures. Similarly, FHA is effective in assessing known hazards and the severity of potential system failures. However, they often rely on expert judgment, introducing subjectivity, and are less equipped to handle non-nominal fault propagation or the effects of multiple component interactions. While FMEA and FHA provide formalized methods to evaluate failure likelihood and event severity, their limitations in modeling complex interdependencies can undermine their effectiveness in modern, highly integrated systems. Though FFIP lacks a formalized risk magnitude assessment process, its systems-level view of failure flows and emergent risks makes it a valuable tool for early-stage SA in complex system designs. One could argue that FHA and FMEA remain popular because they are well-understood, have well-established standards, and serve specific use cases, such as component- and functional-level failure analysis.

Second, we analyzed the current efforts to integrate the FMEA, FHA, and FFIP into MBSE, and categorized them into four. The model-to-model transformation poses potential risks of data loss during conversion from design models to SA tools. Information cross-checking between models is necessary as the transformation process may overlook subtle nuances between models, leading to incomplete safety assessments if not carefully managed. In contrast, the other three methods capture failure and hazard information directly within MBSE tools at varying levels of detail and with varying manual labor effort which raises questions regarding their ability to support digital transformation. These approaches allow consistency between the system safety information and design model, but it come with the risk of data overcrowding within MBSE tools. As more safety-related data is incorporated into the system model, the design information might become harder to interpret, potentially complicating the model's usability for both design and safety engineers. Despite efforts for a total digital transformation, all MBSE integration approaches we documented still require a significant level of additional human effort. Whether through manual reviews of FMEA reports, verifying model transformations, or updating safety-related attributes in MBSE tools, human supervision is still necessary to ensure accuracy and completeness. From our investigation, it is also evident that the literature predominantly focuses on integrating FMEA into MBSE environments, with comparatively limited attention given to FHA and FFIP.

Future research could explore how to effectively integrate early-stage functional SA with later-phase maintenance and SA techniques, such as system health diagnostics. Another research direction could be creating standardized frameworks that unify the integration of SA methods FFIP within MBSE to streamline SA processes in complex systems, and we contend that this could be a promising research area given the proliferation of generative intelligence through standardized ontologies. The text-based foundation of SysML v2 could empower this research trend. Additionally, novel data management strategies are needed to prevent overcrowding in MBSE models, ensuring that safety-related information is represented efficiently. Lastly, exploring how to streamline human intervention in the SA integration with MBSE could be another potential direction of future work.